\def\astroph{1}

\ifnum\astroph=1
\documentstyle[aas2pp4,twoside,natbib]{article}
\else
\fi

\input epsf
\def\s2n{S^{\prime}/N}

\begin{document}
\title{The Stellar IMF from Turbulent Fragmentation}

\author{Paolo Padoan\footnote{padoan@ted.jpl.nasa.gov}}
\affil{Jet Propulsion Laboratory, California Institute of Technology,
MS 169-506, 4800 Oak Grove Drive, Pasadena, CA 91109}
\author{{\AA}ke Nordlund\footnote{aake@astro.ku.dk}}
\affil{Astronomical Observatory and Theoretical Astrophysics Center, Juliane
Maries Vej 30, DK-2100 Copenhagen, Denmark}

\begin{abstract}

The morphology and kinematics of molecular clouds (MCs) are best explained
as the consequence of super--sonic turbulence. Super--sonic turbulence
fragments MCs into dense sheets, filaments and cores and large low density
``voids'', via the action of highly radiative shocks. We refer to this
process as {\it turbulent fragmentation}.

In this work we derive the mass distribution of gravitationally
unstable cores generated by the process of turbulent
fragmentation.  The mass distribution above one solar mass depends primarily on the
power spectrum of the turbulent flow and on the jump conditions for isothermal
shocks in a magnetized gas. For a power spectrum index $\beta=1.74$, consistent
with Larson's velocity dispersion--size relation as well as with new numerical
and analytic results on super--sonic turbulence, we obtain a power law mass distribution
of dense cores with a slope equal to $3/(4-\beta) = 1.33$, consistent with
the slope of the stellar IMF. Below one  solar mass, the mass distribution flattens
and turns  around at a fraction of  a solar mass, as  observed for the
stellar IMF in a number  of stellar clusters, because only the densest
cores  are gravitationally  unstable.   The mass  distribution at  low
masses  is  determined by  the  probability  distribution  of the  gas
density, which is  known to  be approximately Log--Normal for an isothermal
turbulent  gas.   The intermittent  nature  of  the turbulent  density
distribution is thus responsible for  the existence of a significant number
of small collapsing cores, even of sub--stellar mass.

Since  turbulent  fragmentation  is  unavoidable  in  super--sonically
turbulent molecular clouds, and given the success of the present model
in predicting the observed shape  of the stellar IMF, we conclude that
turbulent fragmentation is essential to the origin of the stellar IMF.

\end{abstract}


\section{Introduction}

The process of star formation, particularly the origin of the stellar initial
mass function (IMF), is a fundamental problem in astrophysics. Photometric
properties and chemical evolution of galaxies depend on their stellar
content. The process of galaxy formation cannot be described
independently of the process of star formation, since galaxies are partly made of
stars. Stars are also an important energy source for the interstellar medium
of galaxies.

Stars are formed in molecular clouds (MCs), which
have been the focus of the research on star formation for more than
two decades. Currently, there is no generally accepted theory of star formation,
capable of predicting the star formation rate and the stellar IMF based on the
physical properties of MCs. This is hardly surprising, since turbulent motions
are ubiquitously observed in MCs, and the physics of turbulence is
poorly understood, due to the great mathematical complexity of the fluid
equations. Magnetic field, self--gravity and high Mach numbers further
increase the complexity.

The steady growth of computer performance has now
made large three--dimensional numerical simulations of super--sonic
magneto-hydrodynamic (MHD) turbulence feasible (Padoan \& Nordlund 1997, 1999;
Stone, Ostriker \& Gammie 1998; Mac Low et al.\ 1998; Padoan, Zweibel \&
Nordlund 2000; Klessen, Heitsch \& Mac Low 2000; Mac Low \& Ossenkopf 2000;
Ostriker, Stone \& Gammie 2000; Heitsch, Mac Low \& Klessen 2000;
Padoan et al.\ 2001a,b). Comparisons of numerical experiments with
observational data have shown that super--sonic
turbulence can explain the morphology and kinematics of MCs,
and the formation of dense cores, provided that the motions are
also super-Alfv{\'e}nic
(Padoan, Jones \& Nordlund 1997; Padoan et al.\ 1998;
Padoan \& Nordlund 1997, 1999; Padoan et al.\ 1999; Padoan, Rosolowsky \& Goodman 2001;
Padoan et al.\ 2001a, b). We refer to this process of
formation of dense cores in MCs by super--sonic turbulence as {\it turbulent fragmentation}.
Since protostars evolve from the collapse of gravitationally unstable cores in MCs,
even the stellar IMF could then be the result of turbulent fragmentation,
with the power law shape of the IMF ultimately being the consequence
of the self--similar nature of turbulence.
\nocite{MacLow_Puebla98} \nocite{Stone+98} \nocite{Kimura+Tosa93}
\nocite{Lee+91} \nocite{Vazquez-Semadeni+96} \nocite{Klessen+2000}
\nocite{Padoan+97ext} \nocite{Padoan+98per} \nocite{Padoan+Nordlund98MHD}
\nocite{Padoan+98cat} \nocite{Padoan+Nordlund97MHD} \nocite{Ostriker+99}
\nocite{Passot+95} \nocite{MacLow+98} \nocite{MacLow99}
\nocite{Padoan+2001frag}
\nocite{MacLow+Ossenkopf2000} \nocite{Ostriker+2000} \nocite{Heitsch+2000}
\nocite{Padoan+2000ad}

In this paper we do not use these increasingly sophisticated
numerical simulations directly, but instead develop an analytic
model.  The assumptions of the model are {\em inspired} by the
qualitative properties of the numerical model, but the current work
does not {\em depend} on any particular set of numerical models or
results.

Previous analytic models by Larson (1992), Henriksen (1986, 1991) and Elmegreen
(1993, 1997, 1999, 2000b) have derived the stellar IMF on the basis of the
self--similar structure of MCs. Larson, assuming one dimensional
accretion, predicted a rather steep IMF slope, equal to the MC fractal
dimension, while Henriksen found the IMF slope to depend on both the
MC fractal dimension and the relation between density and linear size
for structures inside MCs.
Elmegreen pointed out that the IMF that results from random sampling of a
self--similar cloud has an exponent $x=1$ (Salpeter $x=1.35$), independent
of the cloud fractal dimension. In Elmegreen's model the IMF is steeper
than $x=1$ because the random sampling rate is assumed to be proportional to
the square root of density, and because of ``mass competition''.

\nocite{Larson92}  \nocite{Henriksen86}  \nocite{Henriksen91}
\nocite{Elmegreen97imf} \nocite{Elmegreen99imf}  \nocite{Elmegreen2000imf}

These models are based on assumptions about the cloud geometry
justified by the apparent fractal structure
of MCs (Beech 1987; Bazell \& D\'{e}sert 1988, Scalo 1990;
Dickman, Horvath \& Margulis 1990; Falgarone, Phillips \& Walker 1991;
Zimmermann, Stutzki \& Winnewisser 1992; Henriksen 1991;
Hetem \& Lepine 1993; Vogelaar \& Wakker 1994; Elmegreen \& Falgarone 1996),
but the processes responsible for generating the assumed geometry
are not discussed in detail.

\nocite{Beech87}  \nocite{Bazell+Desert88}  \nocite{Scalo90}
\nocite{Dickman+90}  \nocite{Falgarone+91} \nocite{Zimmermann+92}
\nocite{Henriksen91} \nocite{Hetem+Lepine93} \nocite{Vogelaar+Wakker94}
\nocite{Elmegreen+Falgarone96}

In a previous attempt to relate the stellar IMF to the physical properties
of super--sonic turbulence (Padoan, Nordlund \& Jones 1997), we obtained the
distribution of the local Jeans' mass in super--sonic isothermal turbulence,
from the probability distribution of the gas density. That work also provides
a prediction for the stellar IMF, by identifying each local Jeans' mass
with a protostar. Although the prediction of the lower mass cutoff and the low
mass portion of the IMF might be roughly correct, this model under--estimates
the number of massive stars relative to low mass stars (as any other Log-Normal
IMF). The main reason, as pointed out by Scalo et al. (1998), is the unphysical
assumption that the most massive stars originate from gas at relatively low
density including only a small fraction of the total mass.

\nocite{Padoan+97imf}  \nocite{Scalo+98}

In the present work we consider specific properties of MC turbulence
and their consequences for the formation of protostellar cores.
In particular, we  i) assume approximate self--similarity of
the super--sonic and super--Alfv{\'e}nic velocity field,
expressed by a power law shape of the power
spectrum of turbulence, and ii) assume that the jump conditions for
isothermal MHD shocks determine the scaling of
the typical size of protostellar cores. Based on these fundamental
assumptions, we are  able to derive a  power law mass distribution
of gravitationally unstable  cores, with a slope consistent with
the Salpeter stellar IMF (Salpeter  1955).
In  order  to  derive  the  mass  distribution  of
gravitationally unstable cores  below one solar mass we  also make use
of  the Probability  Density Function  (PDF)  of the  mass density  in
super--sonic turbulence, following Padoan, Nordlund \& Jones (1997).

The  formation of dense  cores and  the relation  between the  size of
cores and the thickness of the postshock gas are discussed in the next
section.  In  \S~3 the power spectrum  of super--sonic
turbulence is discussed.  Results on  the PDF of the gas density in
super--sonic turbulence are summarized in \S~4.  The mass distribution
of dense  cores formed  by the process  of turbulent  fragmentation is
derived in \S~5, and the mass distribution of gravitationally unstable
cores is  computed in  \S~6.  The relevance  of these results  for the
origin of the stellar IMF is  discussed in \S~7 and 8. Conclusions are
summarized in \S~9.

\section{The Origin of Dense Cores in Super--Sonic Turbulence}

The qualitative properties of the local density maxima that form in
super--sonic turbulence follow from first principles and numerical
experiments have confirmed such properties.

Dense cores are formed in roughly isothermal super--sonic turbulent flows
as the densest parts of sheets or filaments of shocked gas.
Their typical size is therefore comparable to the sheet thickness, $\lambda$.
Assuming that the magnetic pressure in the postshock gas exceeds the
thermal pressure (this is the case for a large range of values of the
preshock magnetic field strength, even when the flow is super--Alfv\'{e}nic),
the isothermal shock jump conditions are:
\begin{equation}
\frac{\rho_1}{\rho_0} \approx {\cal M}_a
\label{mhdn}
\end{equation}
\begin{equation}
\frac{\lambda}{L} \approx {\cal M}_a^{-1}
\label{mhdl}
\end{equation}
\begin{equation}
\frac{B_1}{B_0} \approx {\cal M}_a
\label{mhdb}
\end{equation}
where $\rho_0$, $B_0$ and $\rho_1$, $B_1$ are the values of the gas density
and magnetic field strength before and after the shocks, respectively. $L$ is the
linear extension of the gas before the shock (measured in the direction
perpendicular to the shock surface) and $\lambda$ is the thickness of the postshock
gas. ${\cal M}_a$ is the Alfv\'{e}nic Mach number of the shock, that is the ratio
of the flow velocity and the Alfv\'{e}n velocity measured in the preshock gas:
\begin{equation}
{\cal M}_a = \frac{v}{v_a}=\frac{v}{B_0/\sqrt{4\pi\rho_0}}
\label{ma}
\end{equation} \\
Here the relevant magnetic field components are the ones parallel to the shock surface,
because the perpendicular component does not provide pressure support against the
compression and it is not amplified by the compression. Since only the parallel
components of ${\bf B}$ are amplified, the field in the postshock sheets is nearly
parallel to the sheet surface, and elongated in the direction of dense filaments
(real ones, or two--dimensional sections of sheets).
Since dense cores are formed predominantly in corrugations of sheets, the
magnetic field within dense cores can occasionally show a strong curvature,
depending on the orientation relative to the line of sight. This may considerably
complicate the interpretation of dust polarization measurements
(Ward--Thompson et al.\ 2000) and should be taken into account.

\nocite{Ward-Thompson+2000}

\section{The Power Spectrum of Super--Sonic Turbulence}

The power spectrum of turbulence in the inertial range (below
the energy injection scale and above the dissipation scale)
may be assumed to be a power law,
\begin{equation}
E(k)\propto k^{-\beta} ,
\label{spectrum}
\end{equation}
where $k$ is the wave--number, and the spectral index is $\beta\approx5/3$ for
incompressible turbulence (Kolmogorov power spectrum) (Kolmogorov 1941),
and $\beta\approx2$ for pressureless turbulence (Burgers power spectrum)
(Burgers 1974; Gotoh \& Kraichnan 1993).
The $\beta=2$ power spectrum has often been assumed to be the correct power
spectrum of super-sonic turbulence, at least for the compressional component
of the velocity field.

The purpose of the present work is primarily that of establishing
analytically a relation between the power spectrum of turbulence and the
stellar IMF and we only need to assume that the power spectrum is a power
law in the inertial range. Our analytic model does not assume a specific
value of the spectral index, as obtained for example in numerical simulations.
However, the predictions of our analytic model will be tested using the
power spectrum of super--sonic turbulence computed from numerical
simulations or estimated from observational data. For that reason we
briefly summarize in this section some numerical and observational results.

\subsection{The Power Spectrum in Numerical Simulations}

The most detailed study of the power spectrum of numerical compressible
turbulence has been presented by Porter, Pouquet \& Woodward (1992, 1994)
and Porter, Woodward \& Pouquet (1998). From the 1992 paper to the
1998 paper the largest numerical resolution increased from
$256^3$ to $1024^3$. These works are limited to decaying turbulence, with
Mach numbers close to unity initially, and below unity at later times. The
runs are therefore sub--sonic, except for an initial period of time.
A magnetic field is not included.

The velocity field is usually decomposed into its solenoidal ${\bf v}^s$ and
compressional ${\bf v}^c$ components as ${\bf v}={\bf v}^s+{\bf v}^c$
with ${\bf \nabla \cdot v}^s=0$ and ${\bf \nabla \times v}^c=0$. The
velocity Fourier spectrum, $E(k)$, is also separated into its solenoidal and
compressional parts: $E(k)=E^s(k)+E^c(k)$.

Porter, Pouquet \& Woodward (1992) found that the compressional modes have a
power spectrum $E^c(k)\propto k^{-2}$, and the solenoidal modes
$E^s(k)\propto k^{-1}$. In the later works, after the larger numerical
resolution runs were performed, the same authors concluded instead that both
compressional and solenoidal modes develop a Kolmogorov power spectrum,
$E^c(k)\propto E^s(k) \propto k^{-5/3}$, with $E^c/E^s\approx 0.15$
(Porter, Pouquet \& Woodward 1994; Porter, Woodward \& Pouquet 1998).
This conclusion is not strongly supported by the plots of power spectra
presented by the authors, since the largest resolution runs ($512^3$ and $1024^3$)
are consistent with a Kolmogorov power spectrum for the solenoidal modes only over a very
limited range of wave--numbers, approximately $4<k<10$. At larger wave--numbers,
the power spectrum is flatter, approximately $E^s(k)\propto k^{-1}$.
An interpretation of the shallower power spectrum of solenoidal modes at
large wave numbers is provided, in terms of a ``near dissipation range''.
The discrepancy between the steeper (Burgers) power spectrum of compressional
modes in the early $256^3$ runs (Porter, Pouquet \& Woodward 1992) and the
Kolmogorov power spectrum in the latest $256^3$ runs (Porter, Pouquet \&
Woodward 1994; Porter, Woodward \& Pouquet 1998) is not discussed.

We have recently started to perform a large
number  of numerical  experiments with  the purpose  of  computing the
spectral index of the inertial range of driven super--sonic MHD turbulence as
a function of the sonic and Alfv\'{e}nic rms Mach numbers of the flow.
The experiments  use an isothermal equation of  state, uniform initial
density and magnetic fields,  random initial velocity and random large
scale  forcing  (both  solenoidal),  as described  in  previous  works
(Padoan  et al.\  1998; Padoan  \& Nordlund  1999; Padoan,  Zweibel \&
Nordlund 2000).  The results of these new experiments will be reported
and  discussed elsewhere. Results on
power spectra and structure functions
are reported in Boldyrev, Nordlund \& Padoan (2002),
where the scaling relations are also predicted on the basis
of a new analytic model of super--sonic turbulence (Boldyrev 2002). 
A power spectrum
intermediate between the Burgers and the Kolmogorov power spectra
is found, $E(k)\propto k^{-1.74}$.

\subsection{The Power Spectrum from Observational Data}

The observed velocity dispersion--size Larson relation (Larson 1979, 1981),
$\Delta v\propto L^{\alpha}$, should reflect the power spectrum of
turbulence in the interstellar medium ($\alpha=(\beta-1)/2$),
although it is often obtained from a combination of different clouds and
cores inside the same cloud, or even different molecular transitions
(see Goodman et al.\ 1998 for a discussion of the line width--size relation).
Larson finds $\alpha=0.37$ in the range of scales $1<L<1000$~pc (Larson 1979);
and $\alpha=0.38$ in the range of scales $0.1<L<100$~pc (Larson 1981);
Leung, Kutner \& Mead (1982) obtain $\alpha=0.48$ for $0.2<L<4$~pc;
Myers (1983) gets $\alpha=0.5$ for $0.04<L<10$~pc; Sanders, Scoville \&
Solomon (1985) find an unusually large value $\alpha=0.62$ for
$20<L<100$~pc, which they use to rule out any relation between a turbulent
power spectrum and the line width--size relation; Dame et al.\ (1986)
obtain $\alpha=0.5$ for $10<L<150$~pc; finally Falgarone, Puget \& P\'{e}rault
(1992) use a compilation of data from the literature together with their own
new data, in order to sample a very large range of scales, $0.01<L<100$~pc,
and include also a significant number of unbound objects (velocity dispersion
larger than the virial velocity), which are usually not included in earlier
studies. They find a correlation consistent with $\alpha=0.4$, and a large
total scatter of almost one order of magnitude in line width.
The value $\alpha=0.4$ for the exponent of the line width--size relation
corresponds to a power spectrum of turbulence $\propto k^{-1.8}$.

\nocite{Kolmogorov41} \nocite{Burgers74} \nocite{Gotoh+Kraichnan93}
\nocite{Myers83} \nocite{Falgarone+92} \nocite{Leung+82} \nocite{Sanders+85}
\nocite{Dame+86} \nocite{Goodman+98}

Miesch \& Bally (1994) have estimated the power spectrum of turbulence in molecular
clouds by computing the autocorrelation and structure functions of emission line
centroid velocities. Their results correspond to an average exponent for the
line width--velocity relation $\alpha=0.43$, or a power spectrum with
$\beta=1.86$. Previous attempts to measure the power spectrum of turbulence in
molecular clouds with the same method had provided much shallower spectra
(Kleiner \& Dickman 1987; Hobson 1992).

A new method to estimate the power
spectrum of turbulence in molecular clouds has also been recently proposed
by Brunt \& Heyer (2002), using the Principal Component Analysis by
Heyer \& Schloerb (1997). The
method has already been applied to 23 molecular clouds in the outer Galaxy,
and the result is a power spectrum with exponent varying from cloud to cloud,
in the range $1.72<\beta<2.9$, with a typical error of 0.08. The average
exponent is $\beta=2.17\pm 0.31$. This method has
been calibrated using stochastic fields (Stutzki et al.\ 1998), with purely
random phases and no correlation between density and velocity. Correlations
in real turbulent flows are likely to be important, and could affect the
calibration of this method.

\nocite{Miesch+Bally94}  \nocite{Kleiner+Dickman87} \nocite{Hobson92}
\nocite{Heyer+Schloerb97} \nocite{Stutzki+98}

\section{The PDF of Mass Density}

The study  of Probability Density  Functions (PDF) in  turbulent flows
has received increasing  attention over the last few  years.  PDFs can
provide important information complementary to power spectra.
A well known example
of a  combined use  of power spectrum  and PDF, limited to linear density
fluctuations, is  the Press--Schechter
model of the galaxy mass distribution (Press \& Schechter 1974).

A number  of numerical studies have  established that the  PDF of mass
density  in  isothermal turbulent  flows  is  well  approximated by  a
Log--Normal distribution  (V\'{a}zquez-Semadeni 1994; Padoan, Nordlund
\& Jones 1997; Scalo et al. 1998; Passot \& V\'{a}zquez-Semadeni 1998;
Nordlund \&  Padoan 1999; Ostriker,  Gammie \& Stone 1999), which can
be understood analytically (Nordlund \&  Padoan 1999).  A highly
radiative  turbulent flow  develops  a complex  system of  interacting
shocks that are  able to fragment the mass  distribution into a random
network  of  dense  cores,   filaments  and  sheets  and  low  density
``voids'',  with  a  large  density  contrast. The intermittent nature
of the Log--Normal PDF of mass density means that  most of the
mass  concentrates in  a small  fraction of  the total  volume  of the
simulation.

The Log--Normal distribution may be written as:
\begin{eqnarray}
p(\ln n')d\ln
n'=\frac{1}{(2\pi\sigma^{2})^{1/2}}
\nonumber \\
exp\left[-\frac{1}{2}
\left(\frac{\ln n'-\overline{\ln n'}}{\sigma}\right)^{2}
\right]d\ln n'
\label{pdf1}
\end{eqnarray}
where  $n'$ is  the number  density in  units of  the  average density
$n_0$,
\begin{equation}
n'=n/n_0 ,
\label{pdf2}
\end{equation}
the mean of the logarithm of density, $\overline{\ln n'}$, is determined
by the standard deviation of the logarithm of density,
$\sigma$:
\begin{equation}
\overline{\ln n'}=-\frac{\sigma^{2}}{2} ,
\label{pdf3}
\end{equation}
which is  found to be a  function of the  rms Mach number of  the flow
$\cal{M}$:
\begin{equation}
\sigma^{2}=\ln(1+{\cal{M}}^{2} b^2)
\label{pdf4}
\end{equation}
or, for the linear density:
\begin{equation}
\sigma_{\rho}=b {\cal{M}}
\label{pdf5}
\end{equation}
where $b\approx0.5$, from numerical experiments. The standard
deviation of the linear density distribution thus grows linearly with the
rms  Mach number  of the  flow (Nordlund \& Padoan 1999; Ostriker,
Gammie \& Stone 1999).

\section{The Mass Distribution of Dense Cores}
\label{coremass}

In this section we derive the mass distribution of dense cores based
on the two following assumptions: i) The power spectrum of turbulence is
a power law; ii) The typical size of a dense core scales as the
thickness of the postshock gas.

If the typical size of a dense core is comparable to the thickness of
the postshock gas, $\lambda$ (\S~2), its mass $m$ is:
\begin{equation}
m \sim \rho_1\lambda^3 = \rho_0 {\cal M}_a\left(\frac{L}{{\cal M}_a}\right)^3
= \frac{\rho_0 L^3}{{\cal M}_a^2} ,
\label{m}
\end{equation}
where we have used the jump conditions (\ref{mhdn}) and (\ref{mhdl}).
Equation (\ref{m}) shows that on a given scale $L$, the mass of dense
cores is proportional to the total mass available ($\rho_0 L^3$) divided
by the second power of the Mach number on that scale (${\cal M}_a^2$).

Given the power spectrum (\ref{spectrum}), the rms velocity, $\sigma_v$ on
the scale $L$ is
\begin{equation}
\sigma_v\propto L^{\alpha} ,
\label{larson}
\end{equation}
where
\begin{equation}
\alpha=\frac{\beta-1}{2} .
\label{alpha}
\end{equation}
The typical shock velocity on the scale $L$ is therefore $\sigma_v(L)$, and
the shock Mach number is given by (\ref{ma}), where $v$ is replaced by
$\sigma_v(L)$. Substituting this scale dependent expression of ${\cal M}_a$
into (\ref{m}) one obtains
\begin{equation}
m\approx \frac{\rho_0 L_0^3}{{\cal M}_{a,0}^2} \left(\frac{L}{L_0}\right)^{4-\beta} ,
\label{m2}
\end{equation}
where $L_0$ is the (large) scale where the turbulent velocity is
$v_0$ and the rms Mach number is ${\cal M}_{a,0}$.

The smallest scale where significant density fluctuations may be expected
is approximately the scale
where the Mach number is of order unity:
\begin{equation}
m_{min}\approx \frac{\rho_0 L_0^3}{{\cal M}_{a,0}^{6/(\beta-1)}}
\label{mmin}
\end{equation}
In MCs with a mass $M_0=\rho_0 L_0^3=10^4$~M$_{\odot}$ a typical value of the
Mach number is ${\cal M}_{a,0}\sim 10$. We then obtain
$m_{min}\sim 0.0003$~M$_{\odot}$, for $\beta=1.8$.

If $L_0$ in equation (\ref{m2}) is defined as the largest scale of the turbulent
flow (the scale of turbulent energy injection), and we take $L=L_0$, we obtain
an estimate of the
mass of the largest cores formed by turbulent fragmentation,
\begin{equation}
m_{max}\approx \frac{\rho_0 L_0^3}{{\cal M}_{a,0}^2} ,
\label{mmax}
\end{equation}
where ${\cal M}_{a,0}$ is the rms Mach number on the largest turbulent scale.
In MCs with a mass $M_0=\rho_0 L_0^3=10^4$~M$_{\odot}$
and Mach number ${\cal M}_{a,0}=10$, $m_{max}\sim 100$~M$_{\odot}$.

To arrive at the distribution of core masses we first consider the mass
distribution of a completely self--similar case and then consider the
modification arising from the Mach number dependence.

The completely self--similar case leads to equal mass contributions
from each logarithmic interval (IMF slope -1), as shown by Elmegreen
(1997), who also demonstrated that this result does not depend  on the
fractal dimension of the self--similar distribution.
To illustrate this result, and to derive the modification
of it due to the Mach number dependence, it is useful to consider
the setup and interpretation of numerical experiments.

Numerical turbulence experiments are in a certain sense scale--free;
the density is usually rescaled so that the average density
$\langle\rho\rangle=1$ and the size of the box is scaled so that $L=1$.
The only distinguishing, scale dependent properties that remain
after such scalings are the (sonic and Alfv\'{e}nic) Mach numbers;
from Larson's relations (Larson 1981) we expect larger scales to correspond to
larger rms Mach numbers.

\nocite{Larson81}

To recover the completely self--similar result we consider
two experiments with identical initial conditions (also with respect
to Mach numbers and average gas density), but interpreted at different
scales $L_1$, and $L_2 > L_1$. The total mass in the ``large scale''
experiment is obviously $(L_2/L_1)^3$ larger than in the ``small scale''
experiment.
The cores in the large scale experiment would be equal in number,
but heavier by the ratio $(L_2/L_1)^3$ than cores in the small scale
experiment.  On
the other hand, the total number of cores in the small scale experiment
is $(L_2/L_1)^3$ larger than in the large scale experiment, if the same total
mass is used in the two cases (that is if cores from a number $(L_2/L_1)^3$ of
small scale experiments are counted together). The result is therefore a
total number of cores that depends on scale as:
\begin{equation}
N \propto L^{-3} ~,
\label{nl}
\end{equation}
in agreement with Elmegreen's (1997) result.

When the Mach number dependence on scale is taken into account the
result is that the larger scales contribute relatively less massive cores,
because of the scaling relation (\ref{m2}).  We assume that the number of
cores per scale $L$ still scales as $L^{-3}$.  Combining the relations
(\ref{m2}) and (\ref{nl}) we obtain:
\begin{equation}
N(m){\rm d}\ln m\propto m^{-3/(4-\beta)}{\rm d}\ln m .
\label{imf}
\end{equation}
If the spectral  index  is  consistent  with the  observed velocity
dispersion--size  Larson relation  (Larson 1981)  and with  our
numerical and analytical results (Boldyrev, Nordlund \& Padoan 2001),
then $\beta=1.74$ and the mass distribution is
\begin{equation}
N(m){\rm d}\ln m\propto m^{-1.33}{\rm d}\ln m ,
\label{salpeter}
\end{equation}
which is almost identical to the Salpeter stellar IMF (Salpeter 1955).
\nocite{Salpeter55} \nocite{Larson81}

\section{The Mass Distribution of Collapsing Cores}
\label{collapse}

The mass distribution  of dense cores has been  computed assuming that
the  pre--shock density  is  $n_0$ and  the  postshock density  ${\cal
M}_{a} n_0$, where ${\cal M}_{a}$  is scale dependent.  A more precise
computation should include the  effect of the probability distribution
of the value of ${\cal M}_{a}$ at each scale, or the overall effect of
the  statistics  of  the   turbulent  velocity  field,  which  is  the
generation of a Log--Normal PDF of mass density (see \S~4). This is
necessary   to  compute  the   fraction  of   dense  cores   that  are
gravitationally  unstable and  collapse into  protostars,  since dense
cores can be significantly denser than their average density predicted
by the  scaling laws.   While most  of the large  cores will  be dense
enough to collapse, the probability  that small cores are dense enough
to collapse  is determined by the  PDF of mass density. Because of
the  intermittent  nature of  the  Log--Normal  PDF,  even very  small
(sub--stellar)  cores have  a  finite  chance to  be  dense enough  to
collapse.

We write the thermal Jeans' mass as:
\begin{equation}
m_{J}=m_{J,0}\, \left( \frac{n}{n_0} \right) ^{-1/2}
\label{}
\end{equation}
where:
\begin{equation}
m_{J,0}=1.2\,m_{\odot}\left(\frac{T}{10K}\right)^{3/2}
\left(\frac{n_0}{1000 cm^{-3}}\right)^{-1/2}
\label{mj}
\end{equation}
is the Jeans' mass at the mean density $n_0$. The distribution of the Jeans'
mass is  obtained from the  PDF of density assuming constant temperature
as in Padoan,  Nordlund \& Jones (1997):
\begin{eqnarray}
p(m_J)d\ln m_j = 
 \frac{1}{\sqrt{2\pi}\,\sigma/2}
\left(\frac{m_J}{m_{J,0}}\right)^{-2} 
\nonumber \\
exp\left[{-\frac{1}{2}
\left(\frac{\ln m_J-A}{\sigma/2}\right)^2}\right]
d\ln m_J,
\label{}
\end{eqnarray}
where $m_J$ is in solar masses, and:
\begin{equation}
A=\ln m_{J,0}^2-\overline{\ln n'}
\label{}
\end{equation}
The fraction of cores of  mass $m$ with gravitational energy in excess
of their thermal energy is given by the integral of $p(m_J)$ from 0 to
$m$.  The mass distribution of collapsing cores is therefore
\begin{equation}
N(m)d\ln                   m\propto                  m^{-3/(4-\beta)}
\left[\int_0^m{p(m_J)dm_J}\right]\,d\ln m
\label{imfpdf}
\end{equation}

\ifnum\astroph=1

\begin{figure}
\begin{centering}
\epsfxsize=7cm \epsfbox{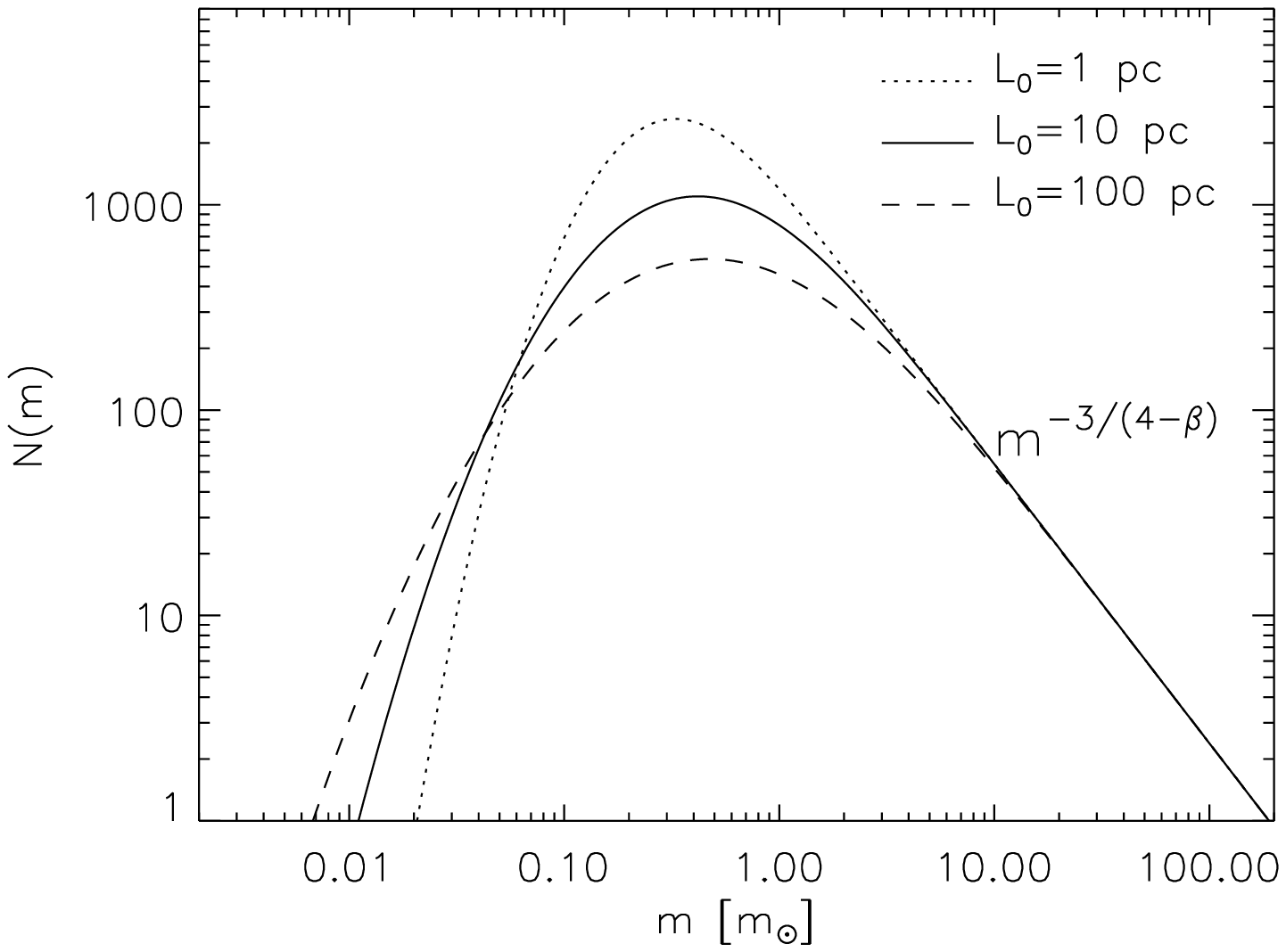}
\epsfxsize=7cm \epsfbox{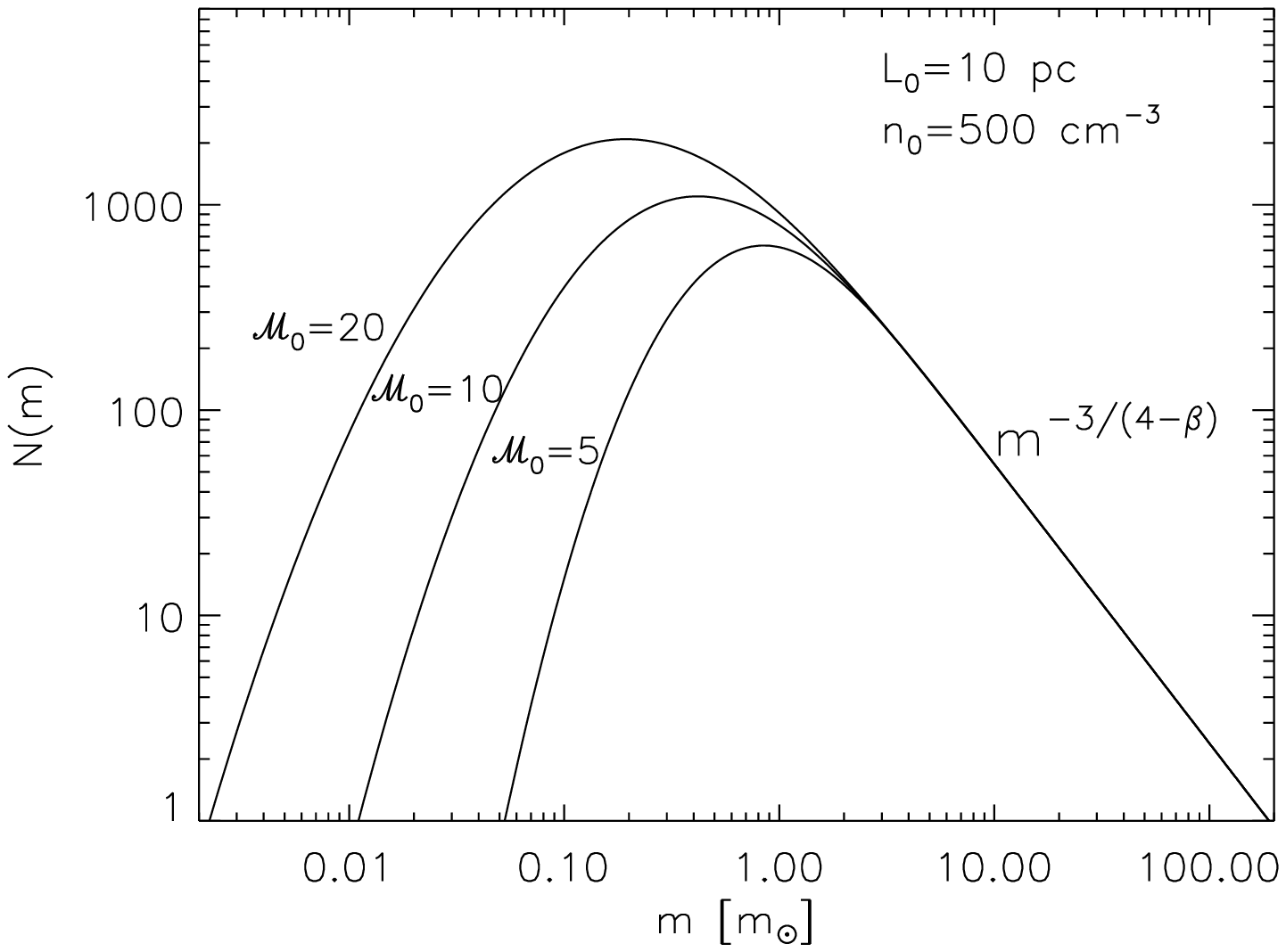}
\epsfxsize=7cm \epsfbox{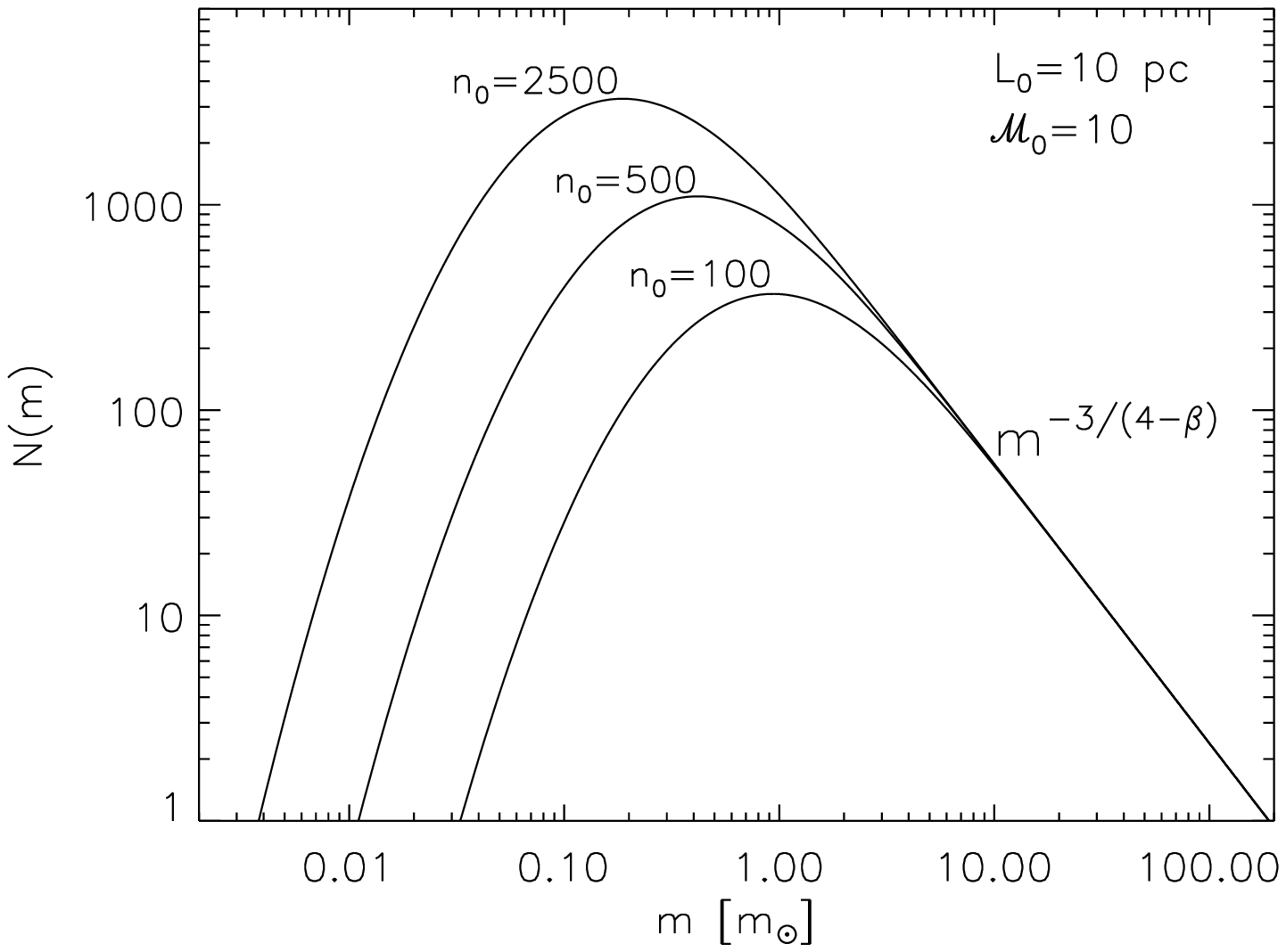}
\end{centering}
\caption[]{%
Mass distributions of gravitationally unstable cores
from  equation (\ref{imfpdf}). Top panel:  Mass distribution for different
values of the largest turbulent scale $L_0$, assuming Larson type
relations (for rescaling $n_0$ and ${\cal  M}_{a,0}$ with $L_0$),
$T_0=10$~K  and  $\beta=1.8$.   Middle  panel:  Mass
distribution  for  different  values  of  ${\cal  M}_{a,0}$,  assuming
$n_0=500$~cm$^{-3}$,  $T_0=10$~K and  $\beta=1.8$. Bottom  panel: Mass
distribution   for  different   values  of   $n_0$,   assuming  ${\cal
M}_{a,0}=10$, $T_0=10$~K and  $\beta=1.8$. The mass distribution peaks at
approximately   $0.4$~m$_{\odot}$,    for   the   values   ${\cal
M}_{a,0}=10$, $n_0=500$~cm$^{-3}$, $T_0=10$~K and $\beta=1.8$, typical of
nearby molecular clouds.
}
\label{fig3}
\end{figure}
\fi

The mass distribution is plotted in Figure~1, for $\beta=1.8$.  In the
top panel the mass distribution is computed for three different values
of the  largest turbulent scale $L_0$, assuming  Larson type relations
(Larson 1981) to rescale $n_0$ and ${\cal  M}_{a,0}$ according to the value
of $L_0$.   The mass distribution  is a power law,  determined by
the power spectrum of turbulence, for masses larger than approximately
1 m$_{\odot}$.  At  smaller  masses the  mass distribution  flattens,
reaches a  maximum at a fraction  of a solar mass,  and then decreases
with  decreasing  stellar mass.   Collapsing  sub--stellar masses  are
found,  thanks  to  the   intermittent  density  distribution  in  the
turbulent  flow.  The  middle and  bottom panel  of Figure~1  show the
dependence of the mass distribution on the rms Mach number of the flow
and on the average gas density respectively.

The magnetic critical mass is derived in the next section. We have not
used  it here  to obtain  the  mass distribution  of collapsing  cores
because  the thermal  Jeans's mass  is  a more  strict condition  for
collapse. The magnetic critical mass depends on the core morphology in
relation to the field geometry and on the magnetic field strength that
correlates with the gas density with a very large scatter (see below).
It is possible therefore that magnetic pressure
support against  the gravitational collapse  affects the shape  of the
mass distribution, but only as a secondary effect.

\section{The Stellar IMF}

Observations show that the stellar IMF is a power law above 1--2 m$_{\odot}$,
with exponent around the Salpeter value $x=1.35$, roughly independent
of environment (Elmegreen 1998, 2000), gradually flattens at smaller
masses, and peaks at approximately 0.2--0.6 m$_{\odot}$ (Hillenbrand 1997;
Bouvier et al.\ 1998; Luhman 1999; Luhman \& Rieke 1999; Luhman 2000;
Luhman et al.\ 2000). The shape of the IMF below 1--2 m$_{\odot}$, and
particularly the relative abundance of brown dwarfs, may depend on the
physical environment (Luhman 2000).

\nocite{Elmegreen_rev98} \nocite{Elmegreen_rev2000}
\nocite{Bouvier+98} \nocite{Luhman99} \nocite{Luhman+Rieke99}
\nocite{Luhman+2000} \nocite{Hillenbrand97}

The scalings discussed above result in a
mass distribution of dense cores
consistent with the stellar IMF for masses larger than 1 m$_{\odot}$,
without invoking a sampling rate proportional to the free fall time,
or ``competition for mass'' as in Elmegreen (1997, 1999).  Two conclusions
are possible; either there are effects in addition to those considered
by Elmegreen, and they all happen to cancel each other, or else additional
effects are not
important in the first place.  In the spirit of Occam, let's consider
the latter possibility.  If, as argued elsewhere by Elmegreen (2000a),
star formation essentially happens in a crossing time, then we may
indeed see only one generation of stars being produced at each scale,
rather than the repeated process implied by scaling with the local
dynamical time.  The picture thus is one where a particular MC forms
as a consequence of the random intersection of counter-streaming,
super--sonic motions (Ballesteros--Paredes, Hartmann \&
V{\'a}zquez--Semadeni 1999; Hartmann, Ballesteros--Paredes \& Bergin 2001), 
internal turbulence creates the distribution of core
masses derived above, and the cores are then grabbed by gravitation to form
one generation of stars.  Energy feedback from stars subsequently disperses
the cloud before the process has time to repeat.

\nocite{Elmegreen2000} \nocite{Ballesteros+99}

In the process envisaged above, turbulent fragmentation is
responsible for creating the core mass distribution, while
gravity is only responsible for the collapse of each protostar.
The flattening and the turn around of the IMF is also easily accounted
for in such a model.
While scale--free turbulence generates a power law mass
distribution down to very small masses, only cores with a gravitational
binding energy in excess of their magnetic and thermal energy can collapse.
The shape  of the stellar IMF is then determined
by the PDF  of gas density, that is by the  probability of small cores
to be  dense enough to  collapse. The mass distribution  of collapsing
cores derived  in the  previous section and  based on  the Log--Normal
PDF of mass density is indeed consistent with the observed IMF.

The scaling of the mass where the IMF peaks can be derived without
a knowledge of the PDF of mass density, using the scaling laws and the
definition of  the critical mass  for collapse. We first  consider the
magnetic critical mass,
\begin{equation}
m_B=m_{B,0}\left(\frac{B}{B_0}\right)^3\left(\frac{n}{n_0}\right)^{-2}
,
\label{mb}
\end{equation}
where $m_{B,0}$ is the magnetic critical mass at the average number
density $n_0$,
\begin{equation}
m_{B,0}=8.3{\rm M}_{\odot} \left(\frac{B_0}{8 {\rm\mu G}}\right)^3
\left(\frac{n_0}{10^3 {\rm cm^{-3}}}\right)^{-2} ,
\label{mb0}
\end{equation}
(McKee et al.\ 1993).
Padoan \& Nordlund (1999) have shown that super--sonic and super--Alfv\'{e}nic
turbulence generates a correlation between gas density and magnetic
field strength, consistent with the observational data. The two most important
properties of such a $B$--$n$ relation are the very large scatter, and the
power law upper envelope ($B\propto n^{0.4}$). More recently, Padoan et al.\ (2001b)
have computed the magnetic field strength in dense cores produced in numerical
simulations of self--gravitating, super--sonic and super--Alfv\'{e}nic turbulence.
They found typical field strength
as a function of column density in agreement with new compilations of
observational samples by Crutcher (1999) and Bourke et al.\ (2000).
Here we adopt the following empirical $B$--$n$ relation consistent with our
previous works:
\begin{equation}
B=B_0 \left(\frac{\rho}{\rho_0}\right)^{0.5} ,
\label{bn}
\end{equation}
where the exponent is $0.5$, and not $0.4$ as reported above, because we now refer
to the average values of $B$ inside bins of $n$, and not to the upper envelope of
the $B$--$n$ relation, as above.  The slight steepening is due to the fact
that the lower envelope of the $B$--$n$ relation is steeper than the upper envelope.

Note that (\ref{bn}) is a statistical relation between the average magnetic field
strength as a function of core density, while Eqs.~(\ref{mhdn}) and (\ref{mhdb}) are
fundamental shock jump relations.  Flows along magnetic field lines are able to
alter the ratio of mass to magnetic flux in regions upstream of shocks, resulting
in the ensemble that (\ref{bn}) characterizes.  Both relations are valid, and enter
into the derivation of the critical mass.

We find the critical mass by imposing $m=m_B$, where $m$ is given by equation
(\ref{m2}), and $m_B$ by equation (\ref{mb}),
\begin{equation}
m_{B,c}= m_{B,0}\left(\frac{\rho L_0^3}{m_{J,0}}\right)^{(\beta-1)/(15-3\beta)}
{\cal M}_{a,0}^{-2/(5-\beta)} ,
\label{mc}
\end{equation}
For $\beta=1.8$, we get
\begin{equation}
m_{B,c}\approx m_{B,0}{\cal M}_{a,0}^{-0.625} .
\label{mcapp}
\end{equation}

\nocite{Crutcher99} \nocite{Bourke+2000}

The critical mass is therefore typically a few times smaller than the
critical mass at the average density. For
${\cal M}_{a,0}=10$, $n_0=10^3$~cm$^{-3}$ and $B_0=8$~$\mu$G, the critical
mass is approximately $2$~m$_{\odot}$.\footnote{This
value of $B_0=8$~$\mu$G provides a normalization of the $B$--$n$ relation consistent
with the results of our super-Alfv\'{e}nic numerical simulations discussed
in Padoan \& Nordlund (1999) and in Padoan et al.\ (2001b).}
The probability that cores smaller
than this mass are larger than their critical mass decreases with decreasing
core mass, which could produce
some flattening of the stellar IMF. Because of the large
scatter in the $B$--$n$ relation the magnetic critical mass does not define
a sharp cut-off in the IMF, but rather a gradual flattening. A sharper
mass scale is defined by the Jeans' mass in the cores,
\begin{equation}
m_{J,c}\approx m_{J,0}{\cal M}_{a,0}^{-2/(5-\beta)} ,
\label{mcj}
\end{equation}
where $m_{J,0}$ is  the Jeans' mass at the  average density defined in
(\ref{mj}).  For  ${\cal M}_{a,0}=10$, $n_0=500$~cm$^{-3}$, $T_0=10$~K
and   $\beta=1.8$,  the   thermal  critical   mass   is  approximately
$0.4$~m$_{\odot}$.   Therefore, using  physical parameters  typical of
nearby molecular clouds,  the present model predicts that  the IMF should
gradually flatten below approximately 2 m$_{\odot}$ and peak in a (logarithmic)
neighborhood of 0.4 m$_{\odot}$, due to increasing thermal pressure support at smaller
masses (cores smaller  than their Jeans' mass are  not included in the
mass distribution).

This  result is  consistent with  the analytic  expression of  the IMF
derived in the  previous section based of the density  PDF.  As can be
seen in  Figure~1, the  IMF peaks at  approximately $0.4$~m$_{\odot}$,
for   ${\cal   M}_{a,0}=10$,   $n_0=500$~cm$^{-3}$,   $T_0=10$~K   and
$\beta=1.8$.

\subsection{The Largest Stellar Mass}

In \S~5 we have estimated the largest mass of dense cores formed
by the process of turbulent fragmentation, given by the expression (\ref{mmax}).
The largest stellar mass should also be of the order of $m_{max}$.
Assuming that the turbulent velocities are of the order of the virial velocities
in the parent molecular cloud, and adopting the Larson relation
$\rho_0\propto L_0^{-1}$ (Larson 1981), we obtain $m_{max}\propto M_{cloud}^{0.5}$
\footnote{According to the Larson relations larger clouds have lower average gas
density than smaller clouds. This does not mean that the most massive stars
are formed in lower density regions. In each cloud stars of different mass
are assumed to form, on average, from pre--shock gas at the mean cloud density.}.
This is close to the empirical relation $m_{max}\propto M_{cloud}^{0.43}$
(Larson 1982), which is found for a cloud sample known to follow the above
size--density relation. It is likely that the true exponent of this relation
is slightly larger than the value found by Larson (1982), because the lifetime
of the most massive stars is comparable to, or shorter than, the lifetime of
their parent molecular clouds, and therefore the probability of observing the
most massive stars decreases with increasing mass.

Elmegreen (1993, 1997) has argued that the largest stellar mass is related to the
mass of the parent cloud for purely statistical reasons: the larger the cloud mass
is, the higher the probability of populating the high mass tail of the IMF.
Such a statistical argument is correct only if the normalization of the
IMF is independent of the total cloud mass.  The scalings derived above show
that this is not the case; larger clouds in general have larger velocities
and therefore form relatively fewer stars of a given mass. This is also
indicated by observations, since it is commonly found that
larger clouds have lower star formation efficiency than smaller clouds.
However, since the star formation efficiency also depends on the low mass
cut-off of the IMF, and since age differences also may enter, it would
have been hard to draw firm conclusions from observations alone.

\nocite{Larson82} \nocite{Elmegreen83}

\section{Discussion}

In Section \ref{coremass} we found that the mass distribution of
dense cores formed by turbulent fragmentation has a power law shape,
with a slope consistent with the stellar IMF at intermediate and large
masses, and in Section \ref{collapse} we found the distribution of
the mass of collapsing cores, which has a shape consistent with the
observed IMF also at small masses.  Three assumptions were essential
for the derivation, and we would here like to emphasize what these
assumptions are, and point out that they are independent of each other.

The first assumption is that approximate selfsimilarity holds for
supersonic turbulence of varying Mach number, to the extent that the
scaling of the number of cores with upstream scale $L$ scales as
$L^{-3}$, also when the Mach number varies with scale $L$. This is a
reasonable assumption because the compressed cores occupy  only a tiny
fraction of the volume where their mass was collected from.  The
second assumption is that the mass of the compressed cores scales as the
cube of the thickness of the shocked gas, which again scales as the
inverse of the Mach number.  This assumption is independent of the
particular density and scale of the upstream flow, and rests only on
the fundamental shock jump condition and on the core shape being
independent of Mach number.  The third assumption is that the Mach
number of the upstream gas depends on the scale $L$ of the upstream
flow as $L^\alpha$, where $\alpha$ is essentially the index in Larson's
(1979, 1981) velocity--size relation.

It is remarkable that, under these assumptions, an IMF slope consistent
with Salpeter's (1955) value is obtained as a direct result of the
observed velocity dispersion--size relation of ISM turbulence and the
fundamental jump conditions for isothermal MHD shocks.
Provided that some general criteria are met, the slope of the stellar
IMF is thus independent of the physical conditions in the star--forming
clouds, as also indicated by the observations (Elmegreen 1998, 2000).

We have also interpreted the gradual flattening of the IMF around 1--2
M$_{\odot}$ and its peak  at approximately 0.3--0.5 M$_{\odot}$ as the
effect  of thermal (possibly also magnetic) support against the  gravitational
collapse. The  mass distribution of collapsing cores,  computed on the
basis of the PDF of mass density and the thermal Jeans' mass, is
consistent  with  the  observed  stellar  IMF,  down  to  sub--stellar
masses.

We have shown that the  IMF at low stellar masses is sensitive
to the  average physical properties of  the star forming  gas, such as
the rms Mach number and  the average gas density, also consistent with
the observations (Luhman 2000).  An even larger number of sub--stellar
objects would be predicted if a significant fraction of them originate
as close companions of more massive stars and later on separate as individual
brown dwarfs. In the present work we do not study the evolution of
density fluctuations smaller than their Jeans' mass within the collapsing
background of an unstable larger core, and therefore this mechanism
for the formation of giant planets and brown dwarfs is not discussed
here.

The assumption behind the result is that the upstream density is
sampled from a Log-Normal density distribution, and that the
distribution of the densities of cores therefore also is approximately
Log-Normal.  An exact Log-Normal density distribution requires exactly
isothermal conditions, but even for non-isothermal conditions the
central part of the density PDF is still approximately Log-Normal
(Scalo et.\ al 1998, Nordlund \& Padoan 1999).

Given the fact that turbulent fragmentation is unavoidable in super--sonic
turbulence, and given the success of the present model
in predicting the correct slope of the stellar IMF,
it is difficult to argue that super--sonic turbulence does not play a dominant
role in the generation of the stellar IMF.
Other processes such as gravitational fragmentation
(Larson 1973; Elmegreen \& Mathieu 1983; Zinnecker 1984), opacity limited
fragmentation (Hoyle 1953; Gaustad 1963; Yoneyama 1972; Suchkov \& Shchekinov 1976;
Low \& Lynden-Bell 1976; Rees 1976; Yoshii \& Saio 1985; Silk 1977a, b),
protostar interactions and coagulation (Nakano 1966; Arny \& Weissman 1973;
Silk \& Takahashi 1979; Bastien 1981; Yoshii \& Saio 1985; Lejeune \& Bastien 1986;
Allen \& Bastien 1995, 1996; Price \& Podsiadlowski 1995; Murray \& Lin 1996),
stellar winds and outflows (Silk 1995; Nakano, Hasegawa \& Norman 1995;
Adams \& Fatuzzo 1996), competitive accretion (Larson 1978; Tohline 1980;
Bonnell et al.\ 1997; Myers 2000) must be relatively unimportant.

This conclusion is supported by a recent computation of the mass distribution of
dense self--gravitating cores in numerical simulations of self--gravitating
super--sonic MHD turbulence (Padoan et al.\ 2001b). The result is a power law mass
distribution consistent with the stellar IMF. Mass distributions of cores from
numerical simulations of super--sonic turbulence, roughly consistent with the stellar IMF,
are also reported by Klessen (2000).

It should be noted that even the formation of molecular clouds, a process that some
authors ascribe to the random collisions of gas streams (e.g. Ballesteros-Paredes,
Hartmann \& V{\'a}zquez-Semadeni 1999; Hartmann, Ballesteros--Paredes \& Bergin 
2001), may be a manifestation of a process
analogous to the one discussed above, just operating on a larger scale.
If motions on larger scales are also characterized by power laws, one would
then expect the distribution of MC masses to obey a Salpeter-like scaling.
However, such scaling would be rather difficult to probe observationally,
due to the fact that MCs are dispersed by the process of star formation
before they are able to collapse as a whole, which prevents them from being
well defined as individual objects.

For almost twenty years estimates of the mass distribution of dense cores in molecular
clouds, based on molecular--line studies, found a shallow power law mass distribution
with a single exponent in the range $0<x<0.7$\footnote{(Myers, Linke \& Benson 1983;
Casoli, Combes \& Gerin 1984; Blitz 1987; Carr 1987; Loren 1989; Stutzkie \& G\"{u}sten 1990;
Lada, Bally \& Stark 1991; Nozawa et al.\ 1991; Tatematsu et al.\ 1993;
Langer, Wilson \& Anderson 1993; Williams \& Blitz 1993; Blitz 1993;
Williams, De Geus \& Blitz 1994; Williams, Blitz \& Stark 1995; Dobashi, Bernard 
\& Fukui 1996;
Onishi et al.\ 1996; Yonekura et al.\ 1997; Kawamura et al.\ 1998).}, with a typical
value $x=0.5$. The main reasons for the shallow mass distribution obtained
in these works are i) the relatively low density traced by the molecular emission
lines normally used ($\sim 10^3$~cm$^{-3}$ for $^{13}$CO, and $\sim 10^4$~cm$^{-3}$
for C$^{18}$O); ii) the limited density range probed by each molecular emission
line; iii) the use of inappropriate ``clump--find'' algorithms; iv) the relatively
low resolution that allows only the selection
of large cores, with typical mass ranging from $\approx 10$~m$_{\odot}$
to hundreds or thousands m$_{\odot}$. Clearly, cores selected in this way cannot be
identified with single protostellar cores, as discussed in Padoan (1995), where it was
predicted that the exponent of the intrinsic mass distribution of protostellar cores
should have been $x>1$, based on the stellar IMF. Recently, Onishi et al.\ (1999) have
obtained a sample of dense cores in the Taurus molecular cloud complex using a higher
density tracer, H$^{13}$CO$^+$, which probes a density of approximately
$n=10^5$~cm$^{-3}$. They found a power law mass distribution with exponent
$x=1.5\pm0.3$, in the mass range between $3.5$~m$_{\odot}$ and $25$~m$_{\odot}$. While
most previous determinations of core mass distributions using molecular--line
maps are affected by the arbitrary definition of an individual core, which is far from
trivial in the hierarchical cloud structure, the dense H$^{13}$CO$^+$ cores found by
Onishi et al.\ (1999) in Taurus are all isolated and therefore unambiguously defined.

\nocite{Padoan95}

Recent dust continuum emission surveys, which also probe relatively high densities
($n=10^5$--$10^6$~cm$^{-3}$), have provided more support to the idea that the stellar IMF
reflects the mass distribution of dense cores.
The mass distribution of dense cores in $\rho$ Ophiuchi (Motte et al.\ 1998; Johnstone
et al.\ 2000), in the Serpens core (Testi \& Sargent 1998) and in Orion (Motte et al.\
2001; Johnstone et al. 2001) are found to be consistent
with the stellar IMF. Motte et al.\ (1998), for example, obtained a power law mass distribution
with exponent $x=1.5$ (in logarithmic units such that Salpeter's exponent is $x=1.35$)
in the range of masses between $0.5$~m$_{\odot}$ and $3$~m$_{\odot}$, and $x=0.5$
in the range between $0.1$~m$_{\odot}$ and $0.5$~m$_{\odot}$. The results by Motte
et al. (1998, 2001) are confirmed by Johnstone et al. (2000, 2001). Testi \& Sargent
(1998) found $x=1.1$, in the range of masses between $0.5$~m$_{\odot}$ and $30$~m$_{\odot}$.

Detailed studies of pre--stellar cores selected in this way could provide
accurate volume density measurements and therefore a test for the density--size
relation implied by our equation (\ref{m}).  Note, however, that because all
quantities involved have wide distribution functions it is important to
distinguish between average or typical values, extreme values, and observed
values.  The latter may often be more representative of extreme values than
of average values.  One example is Larson's density--size relation, which by
its nature tends to refer to the densest cores of a given size rather than
the average cores of that size; many low density cores may exist that are
unobservable, or are ignored.

In the context of our derivation and assumptions one may understand
Larson's density--size relation as the result of combining
Eq.~\ref{mhdn} and \ref{mhdb} and then taking the extreme with respect
to $L$; i.e., the relation represents those extreme cases where roughly
the same initial density gas has been compressed by shocks, perhaps
repeatedly, to various degrees. If each compression is essentially
one--dimensional one retains the approximately inverse relation between
core size and core density.

\nocite{Onishi+99} \nocite{Kawamura+98} \nocite{Myers+83}
\nocite{Casoli+84} \nocite{Blitz87} \nocite{Carr87} \nocite{Loren89}
\nocite{Stutzki+Gusten90} \nocite{Lada+91} \nocite{Nozawa+91}
\nocite{Tatematsu+93} \nocite{Langer+93} \nocite{Williams+Blitz93}
\nocite{Blitz93} \nocite{Williams+94} \nocite{Williams+95}
\nocite{Dobashi+96} \nocite{Onishi+96} \nocite{Yonekura+97}
\nocite{Kawamura+98}

\nocite{Motte+98}  \nocite{Testi+Sargent98} \nocite{Rees76}
\nocite{Yoshii+Saio85} \nocite{Silk77a} \nocite{Silk77b}
\nocite{Nakano66}  \nocite{Arny+Weissman73} \nocite{Silk+Takahashi79}
\nocite{Gaustad63} \nocite{Hoyle53} \nocite{Yoneyama72}
\nocite{Suchkov+Shchekinov76} \nocite{Low+Lynden-Bell76}
\nocite{Allen+Bastien95} \nocite{Allen+Bastien96} \nocite{Lejeune+Bastien86}
\nocite{Bastien81} \nocite{Murray+Lin96} \nocite{Price+Podsiadlowski95}
\nocite{Adams+Fatuzzo96} \nocite{Silk95} \nocite{Nakano+95} \nocite{Myers2000}
\nocite{Larson73} \nocite{Elmegreen+Mathieu83} \nocite{Zinnecker84}

\nocite{Shu+87}

\section{Conclusions}

In conclusion, we have related the stellar IMF to the mass distribution
of dense cores formed by the process of turbulent fragmentation, assuming
that only cores with gravitational energy in excess of their magnetic and
thermal energy can collapse as protostars. Most sub--critical cores disperse
back into the turbulent flow, and are therefore irrelevant for the
process of star formation, as we have argued in other recent works (Padoan
et al.\ 2001a, b). Previous theories of star formation (see Shu, Adams \& Lizano 1987)
assume that stars of small and intermediate mass are formed from sub--critical cores.
In such theories, sub--critical cores are in static equilibrium and
evolve quasi--statically, on the time--scale of ambipolar drift. We have
argued that such a scenario is inconsistent with the turbulent nature of
MCs (Padoan et al.\ 2001a, b).

We have derived the mass distribution of dense cores
generated by the process of turbulent fragmentation, and have found
a power law mass distribution consistent with the Salpeter
stellar IMF,  for stellar masses larger than  1--2 m$_{\odot}$.
This result is a
direct consequence of fundamental physical properties of
super--sonic turbulence in MCs, such as the power spectrum of
turbulence and the jump conditions for isothermal MHD shocks.
We have also  shown that another fundamental  physical property
of turbulent flows, namely the PDF of the gas density, can explain
the shape of the IMF at smaller stellar masses, and the formation
of gravitationally unstable cores of sub--stellar mass.

The main results of this work are:
i) The power law stellar IMF at masses larger than 1--2 m$_{\odot}$ is the
result of the near self--similar nature of inertial range super--sonic
turbulence; ii) The low mass roll-over and cut-off of the IMF is caused by
the combined effects of the thermal support of the smallest cores against
gravitational collapse and the density PDF of super--sonic turbulence; iii)
The mass--scaling of the peak of the IMF may be expressed as a function of
the physical parameters of turbulent star--forming clouds; primarily their
rms turbulent velocity, temperature, magnetic field strength and
average density; iv) For physical  parameters typical of nearby MCs  the
peak  of the  IMF is predicted  to be  around 0.3--0.5 m$_{\odot}$;  v) The
slope  of the  IMF at masses  larger than  1--2 m$_{\odot}$  is determined
by the  inertial range  spectral  index of super--sonic  turbulence and  the
jump  conditions for  isothermal MHD shocks;  vi) For  magnetically
dominated jump conditions, which  are applicable  to typical  molecular
cloud  conditions, the  IMF spectral index is equal  to $3/(4-\beta)$, where
$\beta$ is  the inertial range spectral index  of super--sonic turbulence;
vii) For a value of the inertial range spectral index $\beta=1.74$, consistent
with Larson's relations and with new numerical results, the IMF spectral index is
$x=1.33$, almost identical to Salpeter's slope.

We conclude that the process of turbulent fragmentation is essential to the
origin of the stellar IMF, in support of the thesis that star formation can
be viewed as the main consequence of the dissipation of super--sonic
turbulence in molecular clouds.

\acknowledgements

We thank the anonymous referee for useful comments. We are grateful to
Bruce Elmegreen, Alyssa Goodman, Chris McKee, Phil Myers and Ralph Pudritz
for valuable discussions. This work was performed while PP held a National
Research Council Associateship Award at the Jet Propulsion Laboratory,
California Institute of Technology.
\AA.N.\ acknowledges partial support by the Danish National  Research
Foundation through its establishment of the Theoretical Astrophysics
Center.



\ifnum\astroph=0
\clearpage

\onecolumn

{\bf Figure captions:} \\

{\bf Figure \ref{fig3}:} Mass distributions of gravitationally unstable cores
from  equation (\ref{imfpdf}). Top panel:  Mass distribution for different
values of the largest turbulent scale $L_0$, assuming Larson type
relations (for rescaling $n_0$ and ${\cal  M}_{a,0}$ with $L_0$),
$T_0=10$~K  and  $\beta=1.8$.   Middle  panel:  Mass
distribution  for  different  values  of  ${\cal  M}_{a,0}$,  assuming
$n_0=500$~cm$^{-3}$,  $T_0=10$~K and  $\beta=1.8$. Bottom  panel: Mass
distribution   for  different   values  of   $n_0$,   assuming  ${\cal
M}_{a,0}=10$, $T_0=10$~K and  $\beta=1.8$. The mass distribution peaks at
approximately   $0.4$~m$_{\odot}$,    for   the   values   ${\cal
M}_{a,0}=10$, $n_0=500$~cm$^{-3}$, $T_0=10$~K and $\beta=1.8$, typical of
nearby molecular clouds. \\

\clearpage
\begin{figure}
\begin{centering}
\epsfxsize=9cm \epsfbox{fig1a.eps}
\epsfxsize=9cm \epsfbox{fig1b.eps}
\epsfxsize=9cm \epsfbox{fig1c.eps}
\end{centering}
\caption[]{}
\label{fig3}
\end{figure}

\fi

\end{document}